\documentclass[useAMS,usenatbib]{mn2e}

\title[\emph{MOST} detects corotating bright spots on $\xi$~Per]{\emph{MOST} detects corotating bright spots on the mid-O type giant $\xi$~Persei \thanks{Based on data from the \emph{MOST} satellite, a Canadian Space Agency mission jointly operated by Dynacon Inc., the University of Toronto Institute for Aerospace Studies and the University of British Columbia, with the assistance of the University of Vienna.}\thanks{This paper is dedicated to the memory of Dr. Vincent Fesquet.}}
\author[Ramiaramanantsoa et al.]{Tahina Ramiaramanantsoa$^{1}$\thanks{E-mail:
tahina@astro.umontreal.ca}, Anthony F. J. Moffat$^{1}$, Andr\'{e}-Nicolas Chen\'{e}$^{2,3,4}$, \newauthor Noel D. Richardson$^{1}$, Huib F. Henrichs$^{5}$, S\'{e}bastien Desforges$^{1}$, Victoria Antoci$^{6}$, \newauthor  Jason F. Rowe$^{7}$, Jaymie M. Matthews$^{8}$, Rainer Kuschnig$^{8,9}$, Werner W. Weiss$^{9}$,  \newauthor Dimitar Sasselov$^{10}$, Slavek M. Rucinski$^{11}$ and David B. Guenther$^{12}$
\\\\
$^{1}$D\'{e}partement de physique, Universit\'{e} de Montr\'{e}al, C.P. 6128, Succ. Centre-Ville, Montr\'{e}al, QC, H3C 3J7\\ ~and Centre de Recherche en Astrophysique du Qu\'{e}bec (CRAQ), CANADA\\
$^{2}$Gemini Observatory, Northern Operations Centre, 670 North A'ohoku Place, Hilo, HI96720, USA\\
$^{3}$Departamento de F\'{i}sica y Astronom\'{i}a, Universidad de Valpara\'{i}so, Av. Gran Breta\~na 1111, Playa Ancha, Casilla 5030, CHILE\\
$^{4}$Departamento de Astronom\'{i}a, Universidad de Concepci\'{o}n, Casilla 160-C, CHILE\\
$^{5}$Astronomical Institute "Anton Pannekoek", University of Amsterdam, Science Park 904, 1098 XH Amsterdam, NETHERLANDS\\
$^{6}$Stellar Astrophysics Centre, Dept. of Physics and Astronomy, Aarhus University, Ny Munkegade 120, DK-8000 Aarhus C, DENMARK\\
$^{7}$NASA Ames Research Center, Moffett Field, CA 94035, USA\\
$^{8}$Department of Physics and Astronomy, University of British Columbia, 6224 Agricultural Road, Vancouver, BC V6T 1Z1, CANADA\\
$^{9}$University of Vienna, Institute for Astronomy, T\"urkenschanzstrasse 17, A-1180 Vienna, AUSTRIA\\
$^{10}$Harvard-Smithsonian Center for Astrophysics, 60 Garden Street, Cambridge, MA 02138, USA\\ 
$^{11}$Dept. of Astronomy and Astrophysics, University of Toronto, 50 St George Street, Toronto, ON M5S 3H4, CANADA\\
$^{12}$Institute for Computational Astrophysics, Dept. of Astronomy and Physics, St Mary's University Halifax, NS B3H 3C3, CANADA
}

\usepackage{graphicx}
\begin{document}

\date{ }

\pagerange{\pageref{firstpage}--\pageref{lastpage}} \pubyear{2014}

\maketitle

\label{firstpage}

\begin{abstract}
We have used the \emph{MOST} (Microvariability and Oscillations of STars) microsatellite to obtain four weeks of contiguous high-precision broadband visual photometry of the O7.5III(n)((f)) star $\xi$~Persei in November 2011. This star is well known from previous work to show prominent DACs (Discrete Absorption Components) on time-scales of  about~$2$~d from UV spectroscopy and NRP (Non Radial Pulsation) with one $(l =3)$ p-mode oscillation with a period of $3.5$~h from optical spectroscopy.  Our \emph{MOST}-orbit ($101.4$~min) binned photometry fails to reveal any periodic light variations above the $0.1$~mmag $3$-sigma noise level for periods of hours, while several prominent Fourier peaks emerge at the $1$~mmag level in the two-day period range. These longer-period variations are unlikely due to pulsations, including gravity modes. From our simulations based upon a simple spot model, we deduce that we are seeing the photometric modulation of several co-rotating bright spots on the stellar surface. In our model, the starting times (random) and lifetimes (up to several rotations) vary from one spot to another yet all spots rotate at the same period of $4.18$~d, the best-estimated rotation period of the star. This is the first convincing reported case of co-rotating bright spots on an O star, with important implications for drivers of the DACs (resulting from CIRs - Corotating Interaction Regions) with possible bright-spot generation via a breakout at the surface of a global magnetic field generated by a subsurface convection zone.

\end{abstract}

\begin{keywords}
stars: massive -- stars: O-type -- stars: $\xi$ Persei -- stars: spots  -- technique: photometry
\end{keywords}

\section{Introduction}
\label{sec:intro}

Despite their rarity, luminous hot (massive) stars account for a dominant fraction of the recycling of energy and enriched stellar material back to the ISM which can then form further generations of stars and planets. This occurs both from winds during the whole stellar lifetimes and supernova explosions at the end of their lives. 

Models of the most massive stars near or on the Main Sequence, the O stars, predict that pulsations should occur via the iron-bump $\kappa$ mechanism due to the extension of the $\beta$ Cep instability strip at least to the late-O type stars \citep{1992ApJ...393..272C,1999AcA....49..119P}. This would be indispensable to probe their internal structure.  But such pulsations are not observed photometrically in the majority of O stars \citep{1992MNRAS.254..404B,2010csp..book.....B} which is often the easiest way to probe the variability. On the other hand, spectroscopic surveys for variability in O stars \citep{Fullerton1996ApJS..103..475F} indicate that the vast majority of O stars show photospheric line-profile variations (LPV) that can probably be attributed to pulsation.  In the hottest stars, modes of high-$l$ spherical harmonics are excited in preference to those of low $l$, thus possibly explaining the decline in detectable light variations for stars hotter than B0 \citep{2010csp..book.....B}.  Photometry is most sensitive to very low-$l$ oscillations because of resolution problems and partial cancellation effects at high $l$. Indeed, even if the variations of the partial cancellation factor as a function of $l$ do not behave monotonically, its values decrease at least by a factor of $10$ for $l\geq4$ compared to the dipole mode, and converge towards zero for $l\geq9$, the effect being also more important on odd modes than on even modes \citep[see Equation 6.29 and Figure~6.4~of][]{2010aste.book.....A}.

Strange-mode pulsations are also predicted to occur in the upper part of the Hertzsprung-Russell diagram \citep{1993MNRAS.264...50K,2009CoAst.158..252G}, but it seems that these modes mainly concern stars close to the Humphreys-Davidson limit and observations have not established yet that strange modes are present in O stars.

Most of the few known pulsating O stars are late-O type stars \citep[see][]{1999LNP...523..304H,2005ApJ...623L.145W,2007A&A...463..243D,2008A&A...477..917P,2008A&A...487..659R,2014arXiv1402.6551H}, two of them having the best-established oscillation periods: $\zeta$~Ophiuchi, the best studied of all the O stars by \emph{MOST} (Microvariability and Oscillations of STars) photometry \citep{2005ApJ...623L.145W,2014arXiv1402.6551H}, and HD 93521 \citep{2008A&A...487..659R}, both O9.5V. The $\zeta$ Oph \emph{MOST} light curve is dominated by non-radial pulsations (NRP) in at least a dozen frequencies in the range $1-10~$d$^{-1}$, with dominating period $P = 4.6$~h and amplitudes reaching  $7$~mmag. $\zeta$ Oph and HD 93521 are also amongst the most rapidly rotating stars known, with $v\sin i\simeq400$~km s$^{-1}$ for $\zeta$ Oph and $v\sin i\simeq390$~km s$^{-1}$ for HD 93521. So even if they show low-amplitude $\beta$ Cep-like pulsations, they should be considered as exceptional members of the so-called class of $\beta$ Cep stars \citep{2005ApJS..158..193S}.

Two hotter, mid-O type stars are of particular interest: $\xi$~Persei [O7.5III(n)((f))] and $\lambda$ Cephei [O6I(n)fp], which exhibit short-period LPV, implying NRP with $l = 3-5$ based on high-precision time-dependent spectroscopic monitoring \citep{1999A&A...345..172D}. $\lambda$~Cep is unfortunately not observable by \emph{MOST}.

The COnvection ROtation and planetary Transits satellite \citep[\emph{CoRoT};][]{2006ESASP1306...33B,2009A&A...506..411A} has also been used to study variability in six O stars in a (relatively short for \emph{CoRoT}) run targeting O stars in the young cluster NGC 2244 and its surrounding association Mon OB2 \citep{2011A&A...533A...4B,2010AN....331.1065D,2011A&A...525A.101M,2011A&A...527A.112B}.  These data reveal diverse and highly uncertain origins for the variability, ranging from possible pulsations for those (cooler O) stars closest to the $\beta$ Cep strip (dominated by early-type B stars) to intrinsic red noise mainly for the hotter O stars, with rotation also possible in some cases.  

With regard to their winds, detailed examination of P Cygni absorption troughs of resonant UV lines frequently used to determine mass-loss rates and terminal wind speeds of hot luminous stars, reveals time-variable structures called Discrete Absorption Components (DACs). From their \emph{IUE} (International Ultraviolet Explorer) snapshot survey of $203$ targets, \citet{1989ApJS...69..527H} found that Narrow Absorption Components (NACs) in the P Cygni absorption troughs of unsaturated UV resonance lines are virtually universal among O stars. Subsequent time series observations established that NACs are the end states of propagating DACs, thus implying that DACs are also ubiquitous among O-type stars. More extended data sets of individual stars \citep[e.g.][]{1995ApJ...452L..53M,1995ApJ...452..842M,1995ApJ...452L..65H,1995ApJ...452L..61P,1996A&AS..116..257K}, led to the conclusion that DACs can start out relatively close to the stellar surface as broad absorptions, which accelerate to ever narrower features at higher velocities, asymptotically approaching the terminal wind speed of typically $\sim2000$~km s$^{-1}$.  At any given epoch for an average O star there are on average two dominating DACs per rotation cycle \citep{1999A&A...344..231K}. DACs tend to repeat on a rotation period, yet do not preserve coherency over time scales greater than a few rotations. In the case of $\xi$~Per, \citet{1994A&A...285..565H} used high-time-resolution \emph{IUE} UV spectroscopy of this star to conclude that the variability of its wind takes place in a wide velocity range and most importantly that the observed DACs start out at very low velocities. This was later confirmed by simultaneous H$\alpha$ (close to the stellar surface) and UV wind (far out) observations of $\xi$~Per by~\citet{2001A&A...368..601D} who showed that the DACs are really tied to a region very close to the photosphere.

Concerning the origin of DACs, \citet{1984ApJ...283..303M} suggested that they probably emerge from Corotating Interaction Regions (CIRs) in the stellar wind. These large scale structures are observed and well studied in the solar corona \citep{1972NASSP.308..393H}, where regions of open magnetic field accelerate local parcels of wind plasma ultimately to higher speeds, and the interaction of these streams with the ambient wind, combined with rotation, leads to corotating spiral-like wind perturbations. \citet{1984ApJ...283..303M} extended this paradigm to the case of hot stars and the origin of DACs was associated with CIR compressions within the stellar wind.

\begin{figure*}
\includegraphics[width=16.8cm]{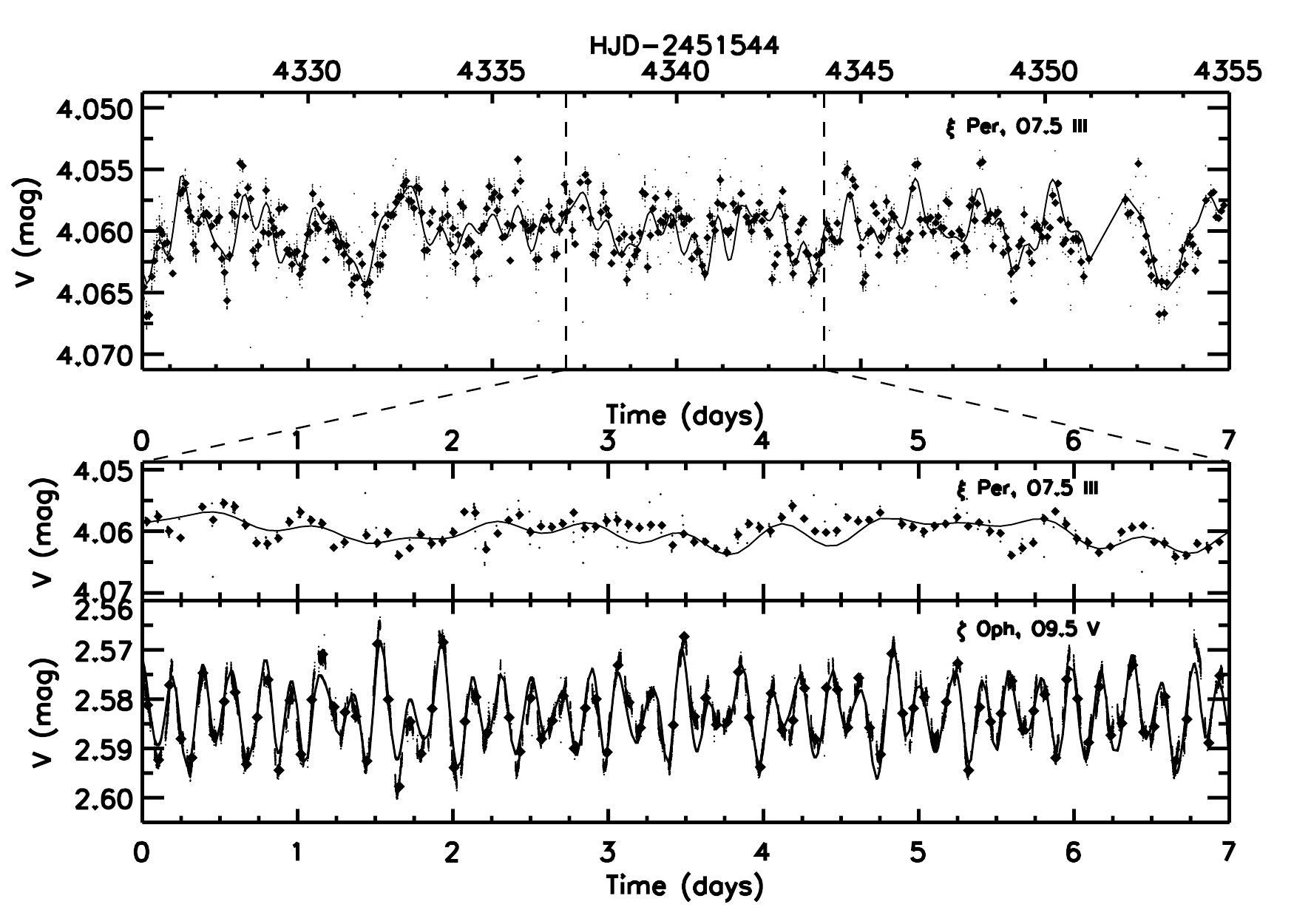}
 \caption{\emph{MOST} photometry. The diamonds are orbital means, whereas the small points represent raw data. \textbf{Top:} Light curve of $\xi$~Per obtained with four contiguous weeks of \emph{MOST} photometry in November 2011. The continuous curve is a fit with the twelve frequencies listed in Table~\ref{tab:period4MOSTlc}. \textbf{Middle:} 7-day sub-sample of the above light curve. \textbf{Bottom:} 7-day sub-sample of the \emph{MOST} light curve of the O9.5 dwarf $\zeta$ Oph for which \citet{2005ApJ...623L.145W} reported the detection of short-period radial and non-radial pulsations. The continuous curve is a 13 frequency fit from Table~1 of \citet{2005ApJ...623L.145W}. Note that the middle and the bottom panels have exactly the same horizontal and vertical scales.}
  \label{fig:xiper_and_zeta_oph}
\end{figure*}

\citet{1994Ap&SS.221..115K} discussed magnetic footprints as the source of DACs in O-type stars. They suggested a low-amplitude, variable dipole magnetic field as the seed of a large-scale wind structure that can be responsible for the emergence of DACs. Later \citet{1996ApJ...462..469C} established models of CIRs in which it turns out that DACs come from extended velocity plateaux forming ahead of the dense CIR compressions. In the simulations of \citet{1996ApJ...462..469C}, bright spots at the stellar surface produce plateau speeds in the wind and absorption features that match with the characteristics of the observed DACs, whereas dark spots generate plateau speeds in the wind that are at $4000-5000$~km s$^{-1}$ (see their~Figure 8a) which are not consistent with the slow-moving observed DACs (Owocki, priv. comm.). The source of the bright spots needed to create the DACs, however, remained a mystery, with NRP or hypothetical magnetic spots remaining as the most likely candidates, with some preference for the latter as the required magnetic field strength would match typical values of the wind confinement parameter $\eta$ for an O star \citep{2008MNRAS.385...97U,2012POBeo..91...13H}. In addition to that, only magnetic spots will co-rotate with the star; NRPs will not be co-rotating, propagating in prograde or retrograde modes.

$\xi$~Per is the brightest single O star in the Northern hemisphere, with the most extensive observational record regarding wind and atmosphere connection, DAC information, pulsation information, and with a magnetic upper limit ($59$~G, model-dependent surface dipole polar field strength for $\xi$~Per - David-Uraz et al., in prep.), but with high-precision photometry desperately missing. Photometry is especially relevant because if DACs arise from co-rotating magnetic spots, the latter should be detectable in optical continuum light, which is the main goal of this investigation.

We have observed $\xi$~Per using \emph{MOST} in an attempt to photometrically reveal its NRP \citep[despite the only known mode so far, a p-mode with $l = 3$, $P = 3.5$~h according to the spectroscopic monitoring of][]{1999A&A...345..172D} and any other variability that might be present (e.g. co-rotating bright spots). It was hoped that this may reveal other pulsation modes, too, which would help clinch the p-modes expected in O stars (and possibly even reveal longer-period g modes from the core region).  We also obtained contemporaneous optical spectra in the range $4000 - 5000$~\AA~at the highest resolution available ($\sim0.4$~\AA~pix$^{-1}$, $\Delta v \sim60$ km s$^{-1}$) and $S/N\sim150$ at the Observatoire du Mont M\'egantic, in order to attempt to match up with the \emph{MOST} photometry and to verify the LPV previously seen in this star \citep{1996A&AS..116..257K,1999A&A...345..172D,2001A&A...368..601D}. Unfortunately due to the insufficient $S/N$ in our spectra and the sparsity of the time coverage, no obvious link could be established between the spectral changes and the simultaneous \emph{MOST} light variations. More extensive spectral coverage will be needed in the future to probe this interesting aspect. Thus, this paper focuses on the outcome of our analysis of the MOST light curve of $\xi$ Per.

\section{Observations}

The photometry presented here was obtained by the \emph{MOST} micro-satellite, which houses a CCD photometer fed by a 15-cm Maksutov telescope through a custom broadband optical filter ($350 - 750$~nm). The satellite's Sun-synchronous polar orbit (period = $101.4$~min, corresponding to a frequency of $14.20$~d$^{-1}$) enables uninterrupted observations of stars in its Continuous Viewing Zone ($-18^o <$  DEC  $< +36^o$) for up to 8 weeks.  A pre-launch summary of the mission is given by \citet{2003PASP..115.1023W} and on-orbit science operations are described by \citet{2004Natur.430...51M}.

$\xi$~Per ($V = 4.06$, RA[2000] =~03:58:57.90, DEC[2000] =~+35:47:27.7) was observed during four contiguous weeks of space-based \emph{MOST} photometry between  2011 November 04 and  2011 December 04 (HJD 2,455,869.5 -- 899.5).  The observations lasted $\sim20$~min of each \emph{MOST} orbit, the remaining orbital time being devoted to two other targets. The data were obtained in Fabry mode and extracted using the technique of \citet{2006MNRAS.367.1417R}.  With such time gaps and no urgent need to look above the \emph{MOST}-orbit Nyquist frequency (or below $P = 2 \times 101$~min = $3.37$~h) we calculated orbital means to create a final light curve (Figure~\ref{fig:xiper_and_zeta_oph}). The standard deviation of each mean point is  $\sim0.3$~mmag.

\section{Results}

\subsection{Rotation period of $\xi$~Per}
\label{subsec:rotation_period}

The rotation period is a crucial parameter in this investigation. Even if the recurrence of the DACs in $\xi$~Per happens at a timescale of $P_{DACs}=2.09$~d, the best estimate of the stellar rotation period is twice that period, i.e. $P_{rot}=4.18$~d \citep[][]{2001A&A...368..601D}.

\begin{figure}
\includegraphics[width=8.4cm]{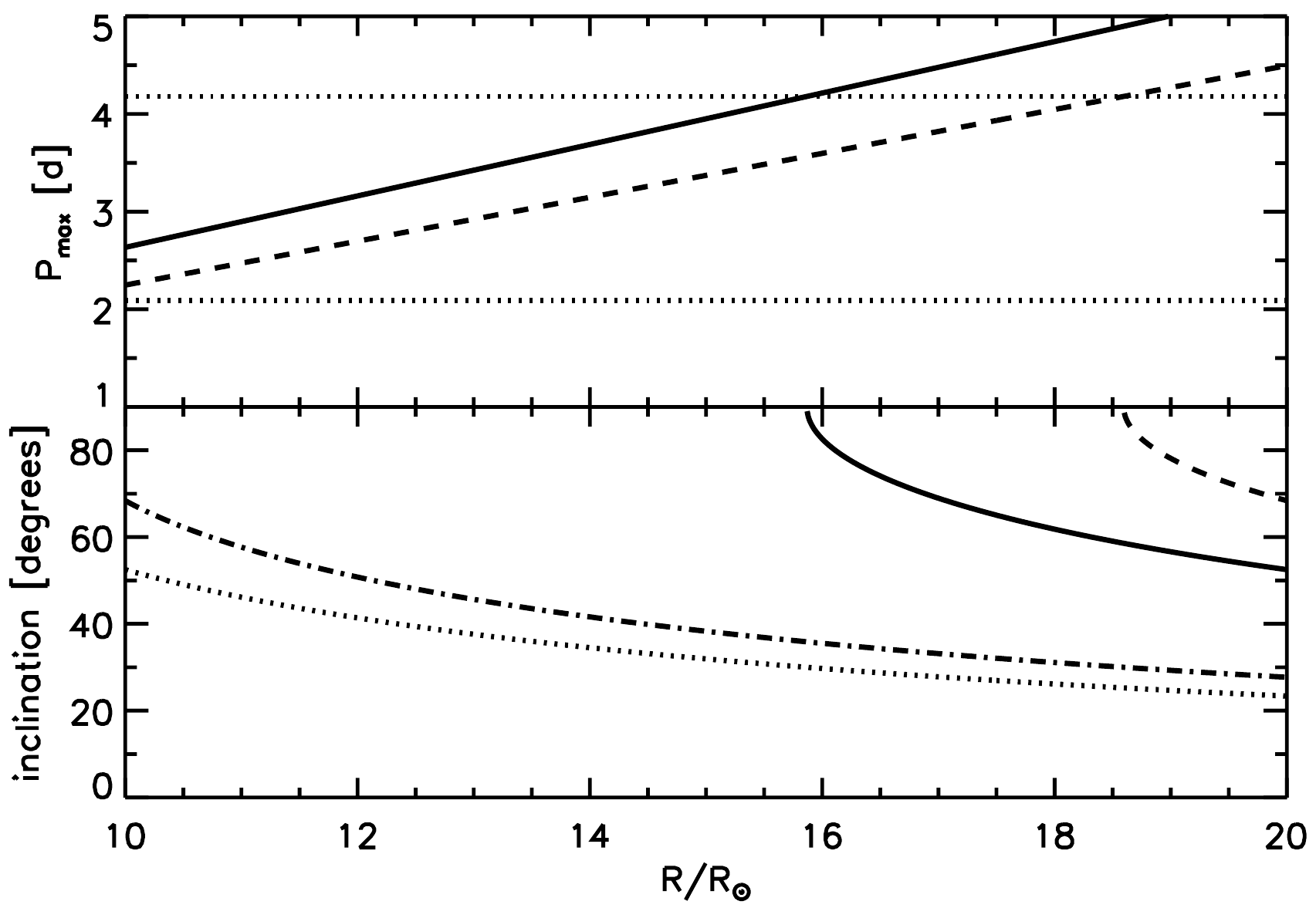}
 \caption{\textbf{Top}: Maximum rotation period as a function of the stellar radius, for $v \sin i=192$~km s$^{-1}$ (solid) and $v \sin i = 225$~km s$^{-1}$ (dashed). The horizontal dotted lines indicate where $P_{max}=2.09$~d and $P_{max}=4.18$~d.  \textbf{Bottom}: Inclination angle as a function of the stellar radius for $v \sin i=192$~km s$^{-1}$ and $P_{rot}=4.18$~d (solid), $v \sin i=192$~km s$^{-1}$ and $P_{rot}=2.09$~d (dotted), $v \sin i=225$~km s$^{-1}$ and $P_{rot}=4.18$~d (dashed), $v \sin i=225$~km s$^{-1}$ and $P_{rot}=2.09$~d (dash dot).}
  \label{fig:xiper_Prot}
\end{figure}

Theoretically, the rotation period is simply related to the stellar radius $R$ and the rotational velocity as $P_{rot}=2\pi R/ v$. The top panel of Figure~\ref{fig:xiper_Prot} shows a plot of the maximum rotation period $P_{max}$ as a function of $R$, for a given value of $v\sin i$. The two extreme values of $v\sin i$ of $\xi$~Per were taken: $192$~km s$^{-1}$ \citep[from][who found $v \sin i=204\pm12$~km s$^{-1}$]{1996ApJ...463..737P} and $225$~km s$^{-1}$ \citep[from][who found $v \sin i=213\pm12$~km s$^{-1}$]{1997MNRAS.284..265H}. For an assumed radius of $14^{+2.1}_{-1.8} R_{\odot}$ \citep[derived by][]{2004A&A...415..349R} and the lowest $v \sin i = 192$~km s$^{-1}$, the period should still be less than $3.6$~d. For higher $v \sin i$ values the period should be even shorter. A $4.18$-day rotation period implies a stellar radius greater than $\sim16 R_{\odot}$. So the adopted radius is critical to accommodate $P_{rot}=4.18$~d. From the bottom panel of Figure~\ref{fig:xiper_Prot}, we can see that if $R\simeq16 R_{\odot}$, a rotation period of $4.18$~d corresponds to an inclination angle of $90^{\circ}$ whereas a rotation period of $2.09$~d yields an inclination angle of $\sim30^{\circ}$. The same reasoning applies for the highest $v \sin i = 225$~km s$^{-1}$: a $4.18$-day rotation period implies a stellar radius greater than $\sim18.5 R_{\odot}$, and when $R\simeq18.5R_{\sun}$, a rotation period of $2.09$~d yields an inclination angle of $\sim30^{\circ}$. However, an inclination angle of the order of $30^{\circ}$ is not consistent with the fact that we see NRP traveling bumps only going from blue to red in the dynamic quotient spectra of $\xi$~Per data \citep[see Figure~2~and~3~of][]{1999A&A...345..172D}.

So this analysis, along with previous spectroscopic analyses by \citet{1999A&A...345..172D} shows that the longer period $P_{rot}=4.18$~d is indeed favoured, and the shorter $2.09$ day period is more or less excluded. This analysis also gives a lower limit on the stellar radius. Our preferred value of $18.5 R_{\odot}$, at a distance of $380$~pc \citep[HIPPARCOS;][]{2007ASSL..350.....V}, implies that the angular diameter of $\xi$~Per is roughly $0.45$~mas, a diameter that would be measurable using interferometry \citep[e.g.][]{2013ApJ...771...40B}.

\subsection{Photometry}
\label{subsec:results_photometry}

\begin{figure}
\includegraphics[width=8.4cm]{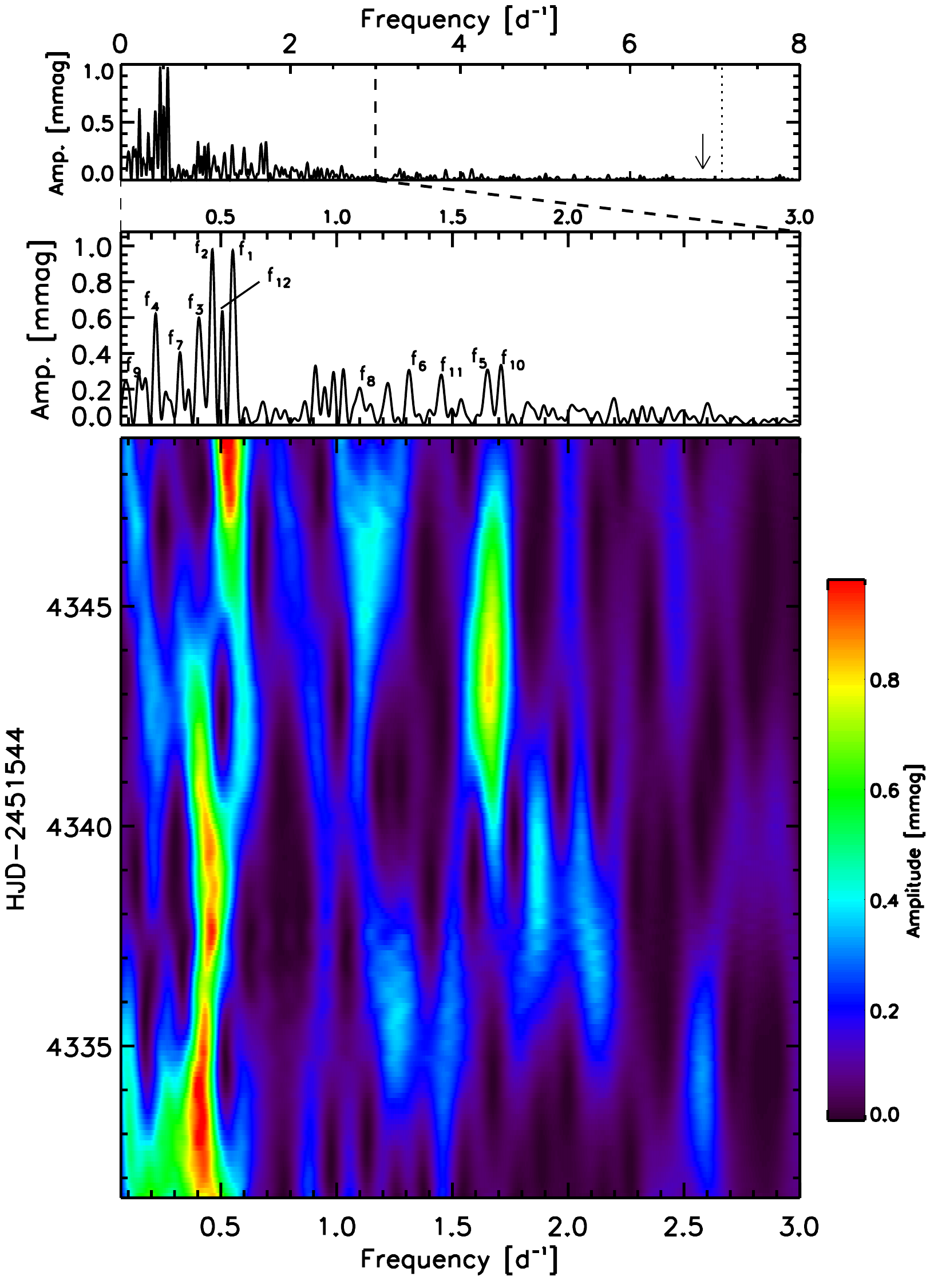}
 \caption{\textbf{Top}: Lomb-Scargle periodogram of the full binned \emph{MOST} light curve of $\xi$~Per, out to and beyond the Nyquist frequency shown as a vertical dotted line at $7.08$~c d$^{-1}$. The arrow indicates that there is no peak at $f = 6.86$~d$^{-1}$ ($P = 3.5$~h). \textbf{Middle}: Zoom on the frequencies below $3$~d$^{-1}$. The twelve frequencies from Table~\ref{tab:period4MOSTlc} are indicated. \textbf{Bottom}: Time-frequency diagram corresponding to the middle panel.}
  \label{fig:xiper_periodogram}
\end{figure}

\begin{table*}
\caption{List of the twelve most significant frequencies from the global Fourier analysis (Period04) of the \emph{MOST} light curve. The starting time for phase values is HJD-2451544 = 4325.547363.}
\centering
\begin{minipage}[c]{\textwidth}
\begin{center}
\begin{tabular}{c c c c c c}
\hline
\hline
 & Frequency [d$^{-1}$] & Period [d] & Amplitude [mmag] & Phase & S/N \\
\hline
$f_1$	&	0.55 $\pm$ 0.003 	&	1.82 $\pm$ 0.01	&	0.90 $\pm$ 0.11 &	0.58 $\pm$ 0.02 &	4.9	\\
$f_2$	&	0.46 $\pm$ 0.006	&	2.17 $\pm$ 0.03	&	0.98 $\pm$ 0.11	&	0.47 $\pm$ 0.02	&	5.2	\\
$f_3$	&	0.41 $\pm$ 0.003	&	2.44 $\pm$ 0.02	&	0.93 $\pm$ 0.11	&	0.64 $\pm$ 0.02	&	5.0	\\
$f_4$	&	0.22 $\pm$ 0.003	&	4.55 $\pm$ 0.06	&	0.70 $\pm$ 0.11	&	0.61 $\pm$ 0.03	&	3.8	\\
$f_5$	&	1.65 $\pm$ 0.004	&	0.61 $\pm$ 0.001	&	0.66 $\pm$ 0.11	&	0.88 $\pm$ 0.03	&	3.6	\\
$f_6$	&	1.31 $\pm$ 0.004	&	0.76 $\pm$ 0.002	&	0.56 $\pm$ 0.11	&	0.96 $\pm$ 0.03	&	3.0	\\
$f_7$	&	0.32 $\pm$ 0.005	&	3.13 $\pm$ 0.05	&	0.50 $\pm$ 0.11	&	0.06 $\pm$ 0.04	&	2.8	\\
$f_8$	&	1.10 $\pm$ 0.004	&	0.91 $\pm$ 0.003	&	0.55 $\pm$ 0.11	&	0.43 $\pm$ 0.02	&	2.9	\\
$f_9$	&	0.09 $\pm$ 0.004	&	11.11 $\pm$ 0.49	&	0.55 $\pm$ 0.11	&	0.11 $\pm$ 0.03	&	2.9	\\
$f_{10}$	&	1.71 $\pm$ 0.004	&	0.58 $\pm$ 0.001	&	0.53 $\pm$ 0.11	&	0.98 $\pm$ 0.04	&	3.0	\\
$f_{11}$	&	1.45 $\pm$ 0.005	&	0.69 $\pm$ 0.002	&	0.50 $\pm$ 0.11	&	0.36 $\pm$ 0.04	&	2.7	\\
$f_{12}$	&	0.50 $\pm$ 0.007	&	2.00 $\pm$ 0.03	&	0.52 $\pm$ 0.11	&	0.92 $\pm$ 0.02	&	2.8	\\ \hline
\end{tabular}
\end{center}
\end{minipage}
\label{tab:period4MOSTlc}
\end{table*}

The outcome of our time-frequency analysis of the \emph{MOST} data is plotted in Figure~\ref{fig:xiper_periodogram}.  On the overall Fourier power spectrum we see a 3-sigma background noise limit of  about $0.1$~mmag at $f = 2$~d$^{-1}$ dropping to $0.03$~mmag at the Nyquist frequency of $7.13$~d$^{-1}$.

Then below $2$~d$^{-1}$ we see a series of about $10$ peaks between $f = 0.8$~d$^{-1}$ [$P=1.25$~d] and $f=1.7$~d$^{-1}$ [$P=0.59$~d] with amplitudes up to $0.3$~mmag, and half a dozen peaks between $f = 0.20$~d$^{-1}$ [$P=5$~d] and $f=0.55$~d$^{-1}$ [$P=1.82$~d] up to $1.0$~mmag. The time-dependent part of Figure~\ref{fig:xiper_periodogram} also shows that power peaks at a given frequency do not last more than a couple of rotations.

These impressions are quantified in Table~\ref{tab:period4MOSTlc}, which lists the $12$ most significant frequencies from the whole data set, using Period04 \citep{2005CoAst.146...53L}.  The three significant peaks ($f_1$, $f_2$, $f_3$) with $S/N > 4.0$ occur in a narrow range from $f = 0.40$~d$^{-1}$ [$P=2.5$~d] to $0.55$~d$^{-1}$ [$P=1.82$~d], which coincides with the dominating DACs frequency at $f = 0.48$~d$^{-1}$ \citep[$P = 2.09$~d :][]{2001A&A...368..601D}; with two nearly equal DACs on average per rotation cycle, this corresponds to the best estimate of the rotation period of $4.18$~d. Two possibly significant frequencies with S/N between $3.0$ and $4.0$ occur at about half and four times this frequency, respectively.  Seven frequencies are marginal ($S/N = 2.5-3.0$) and occur mostly at simple multiples of the primary group of frequencies at $f \sim 0.5$~d$^{-1}$.  Beyond $f = 2$~d$^{-1}$ no outstanding peaks are seen.

In particular, we fail to see the $l = 3$ p-mode from the spectroscopic analysis of \citet{1999A&A...345..172D}, which is not surprising, given that photometry is usually blind to all but the lowest-order pulsations, of which there appear to be none.  The question then arises whether \emph{MOST} may be seeing longer-period (low-order) g-mode pulsations.  While no definitive answer can be given, we do note that we see no recognizable frequency patterns among those peaks that were seen (see~Table~\ref{tab:period4MOSTlc}) that would support this.  Also, fitting the three best frequencies (or even all 12 from Table~\ref{tab:period4MOSTlc}) to the light-curve in Figure~\ref{fig:xiper_and_zeta_oph} leads to a relatively poor match, with significant stochastic residuals.  If pulsations dominated the variability, this kind of curve would match the observed power spectrum very well with negligible residuals, as is the case with the rapidly rotating, pulsating O9.5V star $\zeta$ Ophiuchi \citep{2005ApJ...623L.145W}.  This is not the case for $\xi$~Per, so even longer-period g-mode pulsations do not seem likely.

\begin{figure}
\includegraphics[width=8.4cm]{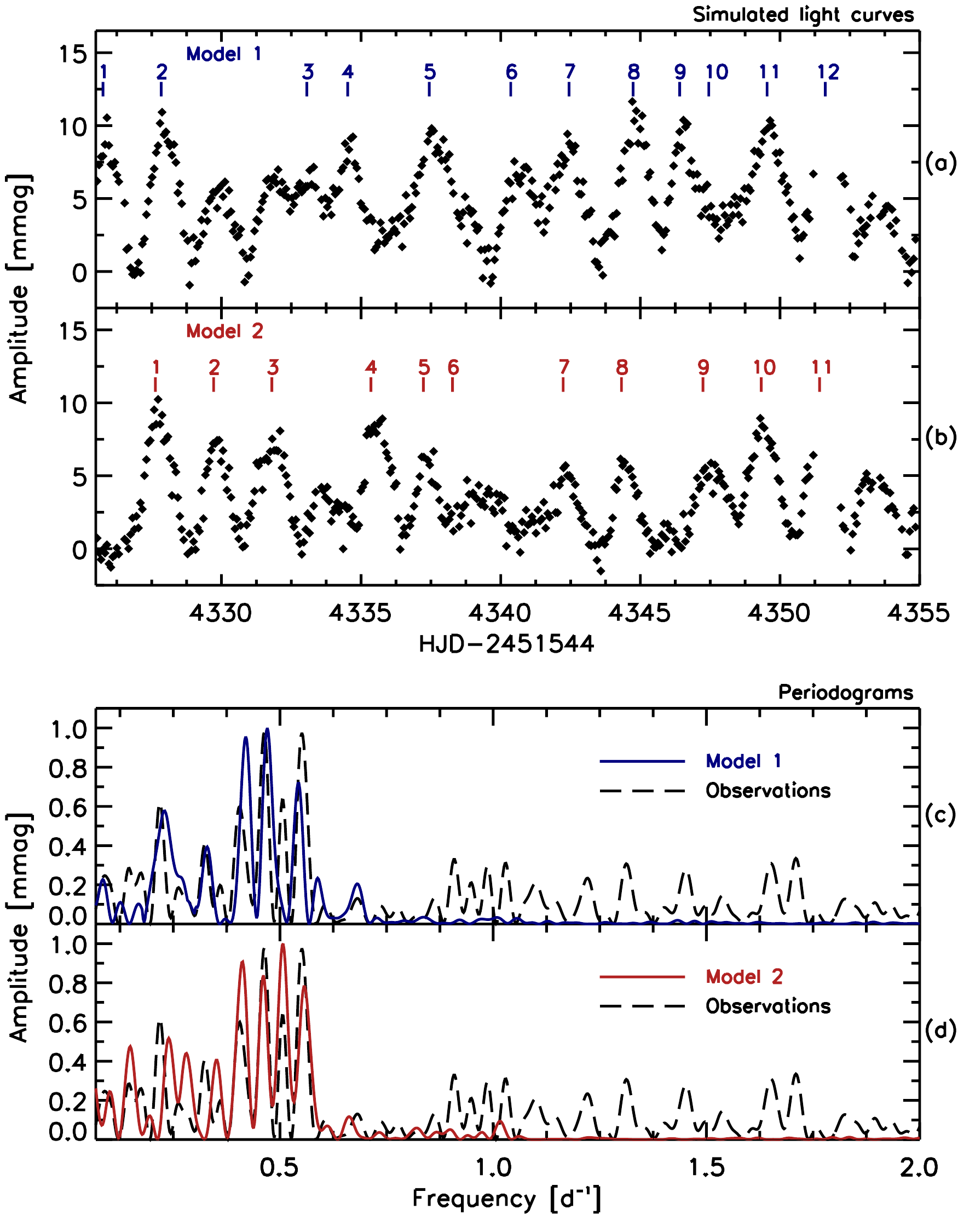}
 \caption{\textbf{Panels (a) and (b)}: Two model light curves with $11-12$ different spots each lasting typically $2-3$ rotations and with random starting times over the same time interval as the actual \emph{MOST} observations. The labeled vertical dashes indicate where each spot reaches its maximum brightness. \textbf{Panels (c) and (d)}: Periodograms of the spot models (blue and red solid curves for model 1 and model 2 respectively) compared to the periodogram of the \emph{MOST} light curve of $\xi$~Per (black, dashed). The power peaks beyond $f = 0.8$~d$^{-1}$ in the observed data are marginally significant and might be accounted for by harmonics or aliases of the main peaks around $0.5$~d$^{-1}$ or by adding more smaller spots.}
  \label{fig:simulSpots_periodogramComp}
\end{figure}

\begin{table*}
\caption{Parameters for each spot in model 1 and model 2. The labels are those used to identify each single spot in Figure~\ref{fig:simulSpots_periodogramComp}. For each spot having a lifetime $\tau$, $t_{max}$ is the time when the spot reaches its maximum amplitude $A_{max}$.}
\centering
\footnotesize
\begin{minipage}[c]{\textwidth}
\begin{center}
\begin{tabular}{c c c c c c c c}
\hline
\hline
&\multicolumn{3}{c}{Model 1}
&\multicolumn{3}{c}{Model 2}\\
\cline{2-4}
\cline{6-8}\\[-2.0ex]
Spot label	&	$t_{max}$ [HJD-2451544]	&	$A_{max}$ [mmag]	&	$\tau$ [d]	&	&	$t_{max}$ [HJD-2451544]	&	$A_{max}$ [mmag]	&	$\tau$ [d]	\\ \hline
1	&	4325.76	&	9.0	&	11.26	&	&	4327.63	&	9.6	&	12.06	\\
2	&	4327.84	&	10.1	&	12.67	&	&	4329.72	&	7.2	&	9.01	\\
3	&	4333.06	&	5.6	&	7.04		&	&	4331.81	&	1.8	&	2.25	\\
4	&	4334.52	&	7.5	&	9.39		&	&	4335.35	&	7.2	&	9.01	\\
5	&	4337.44	&	7.9	&	9.86		&	&	4337.23	&	6.0	&	7.51	\\
6	&	4340.36	&	6.4	&	7.98		&	&	4338.27	&	2.4	&	3.00	\\
7	&	4342.44	&	7.5	&	9.39		&	&	4342.24	&	4.8	&	6.01	\\
8	&	4344.74	&	8.3	&	10.33	&	&	4344.32	&	6.0	&	7.51	\\
9	&	4346.41	&	5.3	&	6.57		&	&	4347.24	&	5.4	&	6.76	\\
10	&	4347.45	&	3.8	&	4.69		&	&	4349.33	&	8.4	&	10.51	\\
11	&	4349.54	&	7.5	&	9.39		&	&	4351.41	&	7.2	&	9.01	\\
12	&	4351.62	&	11.3	&	14.08	&	&	-	&	-	&	-	\\ \hline
\end{tabular}
\end{center}
\end{minipage}
\label{tab:spotsModellc}
\end{table*}

An alternate hypothesis might be that we are seeing variations due to surface bright spots. But then why several frequencies instead of only one dominating frequency?  One conceivable reason might be due to differentially rotating spots at different latitudes overlapping in time.  However, even if there are spots stretching from the equator to near the pole, differential rotation in massive stars is not enough to explain the spread in frequency of even just the three strongest frequencies in Table~\ref{tab:period4MOSTlc} in the range $f = 0.40 - 0.55$ ($\Delta f/f\simeq 30\%$). Actually, considering the models of fast-rotating early-type stars established by \citet{2013A&A...552A..35E}, the surface differential rotation curves as a function of stellar mass tend to be flat beyond $14M_{\odot}$ (see their Figure~15) and an extrapolation up to $40M_{\odot}$ leads to a difference of at most $10\%$ in rotation periods between the equator and the poles.

Therefore, we attempted to test whether we could reproduce the observed main frequencies with a simple model of spots forming and fading, but always with the same rotation period $P_{rot}=4.18$~d. Panels (a) and (b) of Figure~\ref{fig:simulSpots_periodogramComp} show examples of idealized model light curves with 11-12 spots at the equator with $i = 90^{\circ}$ of varying intensity and duration. In these model light curves, we can identify 11-12 main peaks from rotating spots at independent starting times. Each spot lasts a maximum of $2-3$ rotations. This was chosen to be consistent with the behaviour of the time-dependent part of Figure~\ref{fig:xiper_periodogram}. This is also consistent with theoretical predictions that the lifetime of the magnetic field that could be generated in the sub-surface convective zone of massive stars is the turnover time of this convective layer \citep{2011A&A...534A.140C} which, although a priori independent of rotation, turns out to be of the same order as the rotation period. \citet{2011A&A...534A.140C} also predicted a lower limit of the order of hours for the lifetime of a such magnetic spots in a hot massive star, as well as an upper limit of $\sim50$~years for a $20M_{\odot}$ star and  $\sim4$~years for a $60M_{\odot}$ star. It is worth noticing that compared to the Sun, the spot lifetimes considered here are similar to that of large sunspots and sunspot groups \citep{1964ApNr....9...95R,2004soas.book.....F,2010SoPh..262..299H}.

In our model light curves, each small spot is assumed to follow an intensity law of the form $\cos \theta$, where $\theta$ is the angle subtended by the spot as seen from the center of the star, and depends on the rotational phase. The factor of $\cos \theta$ allows for projection effects. Because the intensity should be zero when the spot disappears behind the star, the analytical expression of this function for the $i$-th spot is then:

\begin{equation}
f^{(i)}(t) = \max \left\{A^{(i)}_{max}\cos \left[ \frac{2\pi}{P_{rot}} \left( t-t^{(i)}_{max} \right) \right] , 0\right\}
\label{eq:spot_basic}
\end{equation}

\noindent where $P_{rot}$ is the rotation period and $t^{(i)}_{max}$ the time at which the spot reaches its maximum amplitude $A^{(i)}_{max}$. This main function is modulated by the growth law and the decay law. In the case of the Sun, sunspots emerge fast on a timescale of $\sim$ hours \citep[see e.g.][]{2003A&ARv..11..153S} and start to decay right after their emergence. The decay law of sunspots was initially thought to be linear \citep[e.g.][]{1963BAICz..14...91B,1988A&A...205..289M} but quadratic laws were also proposed \citep[e.g.][]{1997SoPh..176..249P}. In our simulations, the growth law and the decay law are assumed to be simply linear, resulting in an envelope function of the form:

\begin{equation}
\mathcal{E}^{(i)}\left(t\right) = 
\left\{
\begin{array}{rl}
 {\displaystyle \frac{1}{\tau^{(i)}_{rise}}} \left[t-t^{(i)}_{max}+\tau^{(i)}_{rise}\right]  & \mbox{for } t \leq t^{(i)}_{max} \\[3.0ex]
{\displaystyle -\frac{1}{\tau^{(i)}_{dec}}} \left[t-t^{(i)}_{max}-\tau^{(i)}_{dec}\right]  & \mbox{for } t > t^{(i)}_{max}
\end{array}
\right.
\label{eq:spot_enveloppe}
\end{equation}

\noindent where $\tau^{(i)}_{rise}$ and $\tau^{(i)}_{dec}$ are the \emph{duration} of the growth and the decay phases respectively. The growth time of a spot is taken to be a tenth of the duration of the decay phase, so that these two parameters are not independent and can be both constrained when defining the lifetime $\tau^{(i)}=\tau^{(i)}_{rise}+\tau^{(i)}_{dec}$ of the spot. Also, by analogy to the Gnevishev-Waldmeier relation for the Sun, which states the proportionality between the maximal area $\mathcal{S}_{max}$ of a sunspot group and its lifetime $\tau$ \citep[$\mathcal{S}_{max}/\tau=10$ millionth of the solar hemisphere per day:][]{Gnevyshev1938, 1955QB521.W3.......} and whose validity for individual sunspots was established by \citet{1997SoPh..176..249P}, the ratio $A^{(i)}_{max}/\tau^{(i)}$ is assumed to be constant in our simulations. The constant of proportionality was assessed by assuming that the largest spot that can induce a variation of up to $\sim10$~mmag in the light curve (see Figure~\ref{fig:xiper_and_zeta_oph}) would last three rotations. This yields a constant of proportionality of $0.8$~mmag d$^{-1}$ between $A^{(i)}_{max}$ and $\tau^{(i)}$.

Finally, the synthetic light curve containing $N$ spots is defined by $\mathcal{I}\left(t\right)= \sum_{i=1}^{N}f^{(i)}\left(t\right)\cdot\mathcal{E}^{(i)}\left(t\right)$. We also include white gaussian noise in our simulations. No assumptions were made concerning the exact shape of the spots and their angular extent in two dimensions, assuming that the definitions of $f^{(i)}\left(t\right)$ and the envelope function $\mathcal{E}^{(i)}\left(t\right)$ for each spot are sufficient to study the light variations that they induce in the continuum. Also, it is worth noticing that our simulations do not include limb darkening effects. Table~\ref{tab:spotsModellc} summarizes the parameters ($t^{(i)}_{max}$, $A^{(i)}_{max}$, $\tau^{(i)}$) for each individual spot in the synthetic light curves presented in Figure~\ref{fig:simulSpots_periodogramComp}. Although we do not expect the simulated light curves to match the observed one in any exact way, the power spectra in panels (c) and (d) of Figure~\ref{fig:simulSpots_periodogramComp} reveal four significant power peaks in a similar range as observed for the main frequencies.  The remaining peaks in the simulations are harmonics of the stronger peaks. It should be emphasized that this is not a unique solution; true simulation of the observed light curve is a formidable task, with many parameters (e.g. spot number, intensity-distribution, starting time, duration, decay law, etcÉ) for which we have no real independent constraints. Treating this as an inverse problem would also be impractical, given the complexity and uncertainty of the nature of the spots.
Nevertheless, we are content to have found two solutions which at least qualitatively match what we see; in fact there are many more possible solutions.  Note that the time-frequency analysis shows wandering frequencies, compatible with the idea of a multitude of interfering spots with different start-times and durations but all rotating at the same rate, or nearly so.

\section{Discussion}

Our analysis (Section~\ref{subsec:results_photometry}) shows that the \emph{MOST} photometric variability  is predominately driven by rotating surface bright spot modulations. The largest question is the origin of such spots. If these spots are co-rotating, they must be fixed to the stellar surface like sunspots and thus not be related to NRP. If non co-rotating, the spots may be related to low g-mode NRP, either prograde or retrograde.  The former interpretation (co-rotation) is more consistent with the behaviours of the \emph{MOST} light curve and its periodogram showing that the rotation time-scales match the estimated stellar rotation period of $P_{rot}=4.18$~d. However, at this point it is worth noting that our models which reproduce the Fourier spectrum (see~Figure~\ref{fig:simulSpots_periodogramComp}) involve more than two co-rotating spots per rotation cycle. This may suggest that there could be many more DACs of progressively lower amplitudes in $\xi$~Per, in contrast to the dominant two-DAC per rotation view of $\xi$~Per that has been seen with \emph{IUE} \citep{1999A&A...344..231K}. The detectability of these DACs of lower strengths requires high signal-to-noise, high spectral resolution ultraviolet spectroscopy.
 
In the \emph{MOST} light curve, one might wonder if instead we are seeing the photometric modulations directly of DACs/CIRs and not the bright spots that might cause them. However, for the low density wind of $\xi$~Per \citep[$\dot{M} \leq 1.2\times10^{-6}M_{\odot}$ yr$^{-1}$ including clumping corrections:][]{2006A&A...454..625P}, the stellar wind does not emit continuum light in the optical. A study of the spectral energy distributions of Be stars and evolved massive stars showed that, apart from the cases of extremely strong $\dot{M}$ or disk, the flux excess is not important in the optical \citep{2010PASP..122..379T}. The stellar wind only becomes important at long wavelengths where free-free and bound-free emission processes begin to dominate and cause larger amounts of flux excess. The stellar wind only scatters continuum light from the central star, which then becomes highly diffused.

The connection of the light variations to co-rotating spots is surprising because such hot stars were long thought not to have convection zones except deep in their interior core region where highly temperature-dependent nuclear fusion is going on.  This has changed, however, with the modelling work of \citet{2009A&A...499..279C}, who find for the first time a subsurface convection zone due to partial ionization (PIZ) near $T \sim 170 000$~K of the abundant Fe-like elements.  This zone is capable of driving acoustic and gravity waves, which reach the stellar surface through a radiative skin and in turn could create turbulence in the strong stellar winds from hot luminous stars, which manifest themselves as wind clumps. This PIZ may also be the source of a global magnetic field, winding up toroidally with stochastic buoyancy breakouts at the surface causing co-rotating magnetic bright spots at the surface of the star \citep{2011A&A...534A.140C}.  Such spots would normally occur in close north-south polarity pairs, as on the Sun, and would not be easily detectable via Zeeman splitting with current spectro-polarimeters due to cancelation effects of the close bipolar field. \citet{2013A&A...554A..93K} estimated the detectability of a longitudinal magnetic field in early-type stars by putting random magnetic spots on the stellar surface and using results of the recently completed Magnetism in Massive Stars survey \citep[MiMeS;][]{2011IAUS..272..118W,2012ASPC..464..405W}. Under a number of conditions such magnetic spots are not detectable with the best available instruments (CFHT/ESPaDOnS, TBL/NARVAL, ESO/HARPSpol). Even the global field could be below current detection limits ($59$~G upper limit for $\xi$~Per as mentioned in section~\ref{sec:intro}). However, when these spots (unlike the optically dark spots on the Sun) rotate with the star, they presumably cause some of the photometric (continuum light) modulations that we see in some O stars.

These co-rotating magnetic spots were also proposed to be at the base of the so-called "stellar prominences" that trigger the cyclical variabilities observed in the wind lines of some O stars, e.g. the case of the O6 supergiant $\lambda$ Cep \citep{2013arXiv1310.5264H}. Concerning the latter, \citet{2014arXiv1401.2875U} recently reported an intensive spectroscopic study of this star in search of the two pulsation frequencies found by \citet{1999A&A...345..172D}. Focusing on lines stronger than the He{\footnotesize I} $\lambda$4713 line studied by \citet{1999A&A...345..172D}, their spectra with $S/N$ lower than that of the data used by \citet{1999A&A...345..172D} failed to reveal any stable frequency which might be connected to pulsations.  Given that $\lambda$~Cep and $\xi$~Per share many similar properties, including the presence of strong DACs that do not repeat exactly from one rotation to the next, we suggest that the complex behaviour of the line profile variations reported by \citet{2014arXiv1401.2875U} are the results of rotating spots. These spots come and go on timescales of up to several times the rotation period similar to the spots we observed on $\xi$~Per.

\section{Conclusion and future work}

Some recent observations of high-precision photometric monitoring, especially from space where one can collect large amounts of precise, high-cadence, uninterrupted data, are revealing the presence of rotating spots on the surfaces of hot luminous OB stars \citep[e.g.][]{2010A&A...519A..38D,2013A&A...557A.114A} and their descendant WR stars \citetext{\citealp{2011ApJ...735...34C}; Moffat et al., in prep.}. Most likely we are seeing the detection of co-rotating bright spots at or near the stellar surface. $\xi$~Per is the first clear case of co-rotating bright spots on an O star along with a plausible link to DACs. Because the case of $\xi$~Per is not unique regarding UV DACs, this kind of photometric covariability should be observable and should be similar in other O-type stars showing DACs.

More work is needed to detect co-rotating bright spots in other O stars, preferably using high-precision photometric monitoring from space. Most late-type O stars are still in the $\beta$ Cep instability strip, so their light curves could be dominated by NRP, like that of $\zeta$ Oph \citep{2005ApJ...623L.145W,2014arXiv1402.6551H}. Thus it would be more appropriate to look at early/mid-type O stars. The next generation of nanosatellites for asteroseismology, BRIght Target Explorer (\emph{BRITE-Constellation}; http://www.brite-constellation.at/), consisting of a network of six independent 30-mm space telescopes, is expected to photometrically monitor $36$ O stars with $V < 6$~mag at a precision level of $\sim20$~ppm in Fourier space. Fourteen of these targets are early/mid-type O stars. The mission will then contribute in determining if the co-rotating bright spots phenomenon is universal among O stars. Such bright spots may be the drivers of DAC activity, ubiquitous to all O and maybe also their descendant WR and other types of evolved massive stars.

\section*{Acknowledgments}

We gratefully acknowledge useful conversations with Paul Charbonneau, Nicole St-Louis, Stan Owocki and Alex Fullerton in relation to this project. D.B.G., J.M.M., A.F.J.M., and S.M.R. are supported by NSERC (Canada), with additional support to A.F.J.M. from FQRNT (Qu\'ebec). A.N.C. gratefully acknowledges support from the Chilean Centro de Astrof\'isica FONDAP No. 15010003 and the Chilean Centro de Excelencia en Astrof\'isica y Tecnolog\'ias Afines (CATA). A.N.C. also received support from the Comite Mixto ESO-Gobierno de Chile and GEMINI-CONICYT No. 32110005. V.A. acknowledges the Stellar Astrophysics Centre (SAC) funded by The Danish National Research Foundation. V.A. also received support from the ASTERISK project (ASTERoseismic Investigations with SONG and Kepler) funded by the European Research Council (Grant agreement no.: 267864). R.K. and W.W.W. are supported by the Austrian Space Agency and the Austrian Science Fund. NDR acknowledges his CRAQ (Centre de Recherche en Astrophysique du Qu\'ebec) fellowship.

\bsp

\label{lastpage}

\end{document}